\documentclass[aps,pre,twocolumn,groupedaddress,showpacs]{revtex4}  
\usepackage{graphicx}   
\usepackage{subfigure}   
\usepackage{bm}

\newcommand{\ncl}{$N_{cl}$}  
\newcommand{\nev}{$t_N^{clog}$}  
\newcommand{\nzero}{$N_F$}  
\newcommand{\kcl}{$K_{cl}$}

\newcommand{\tw}{$\tau$}  
  
\newcommand{\dtw}{$P(\tau$)}  
\newcommand{\st}{$\lambda$}  
\newcommand{\lglg}{$\log$-$\log$}  
\newcommand{\szf}{$s$}  
\newcommand{\szp}{$r$}  
\newcommand{\mup}{$\mu$}

\newcommand{\fa}{({\em a})}  
\newcommand{\fb}{({\em b})}  
\newcommand{\fc}{({\em c})}

\newcommand{\inc}{{\em incremental clogging}}  
  
\begin{document}  
  
  
\title{Dust Control in Finite Air Volumes at Zero Gravity - Mean-Field Like Analysis}  
  
\author{T.~R.~Krishna Mohan} 
  \altaffiliation[On leave from ]{CSIR Centre for Mathematical Modelling and Computer Simulation (C-MMACS), Bangalore 560017, India.} 
\author{Surajit Sen}  
\affiliation{Department of Physics, State University of New York, Buffalo, New York 14260-1500}  
\email{kmohan@buffalo.edu, sen@dynamics.physics.buffalo.edu}  
  
\date{\today}  
  
\begin{abstract}

We study a simple 1$D$ model of dust rods, with mean size
$\mu$, passing through a parallel 1$D$ alignment of pores as a 
problem of clogging of a filter by dust grains; \mup\ is
kept less than the pore size, $s$. We assume that the filter is ``sticky",
characterized by some parameter $0 \le \lambda \le 1$, which means that dust
grains slightly smaller in size than $s$ can get trapped in the pores. Our analyses
suggest that the number of clogged pores, $N_{cl}$, grows in time as
\nev\ $\propto N_{cl}^{\nu}$, where $\nu = \nu(\mu,\lambda)$ is a non-universal
exponent that depends upon the dust size distribution and filter
properties. 

 
 

\end{abstract}
  
\maketitle  
Common household dust typically comprises of particulate matter such as carbon particles,   
skin flakes, dust mites, fiber, hair and other irregular shaped objects with length-scales between   
$10^{-2}\mu$m and $10^{2}\mu$m (see, for example, \cite{bro}). Typically, the larger sized hairs/fibers act as nucleation centers   
for agglomeration of fine scale dust and the eventual development of the familiar dust bunnies.  
Presence of dust bunnies reduces the effectiveness of dust filters. We study the problem of control of dust at   
all length-scales to ensure that dust bunny growth becomes   
difficult. In particular, we consider dust control at zero gravity in an air volume that is finite and   
cannot be replenished. Our mean-field like analysis is inspired by   
the problem of air quality maintenance in the International Space Station. The present study has the potential to facilitate the design of filters which can retain air-quality below specified thresholds, for long periods.  
  
Significant work has been done to describe the pressure drop in fluids with   
fine particulates that is flowing through filters with given specifications (see, for example, \cite{wata}). The analyses have been performed typically via rate equations using state variables, akin in spirit to thermodynamical analysis. The resultant {\it blocking equations} reveal the behavior of pressure drop with time as filtrate volumes are modified  and filter criteria are tuned~\cite{hebr}. Limited information appears to be available on specifics such as time-dependent degradation of filters as functions of dust size distributions and filter pore distributions.  
  
The present work focuses on filter clogging as a function of time. We refrain from addressing the dust agglomeration process in this Letter. Fine scale dust is continuously generated in most inhabited environments and naturally agglomerates. Dust mites ($\sim 10^{2}$ $\mu$m), plentiful in most   
environments, tend to feed on smaller length scale dust. Presence of dust mites therefore   
leads to a natural integration of smaller length scale dust into larger length scale dust.  
 
\begin{figure*}  
\centering
\includegraphics[scale=0.80]{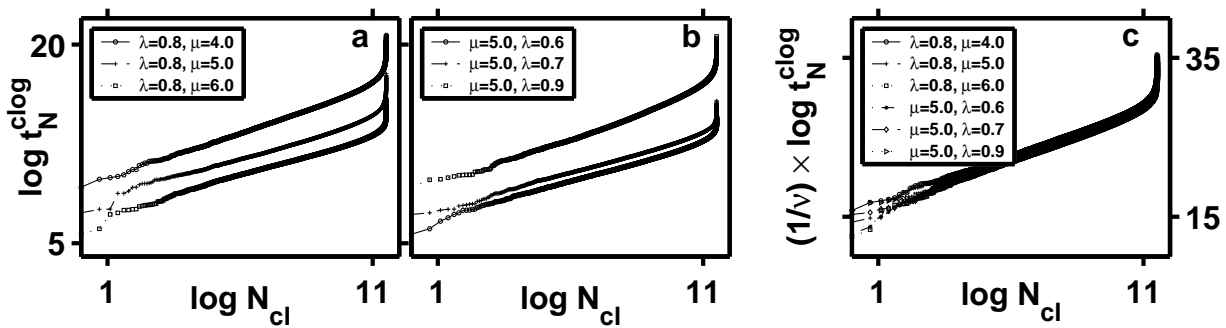}  
\caption{In Fig.~\ref{scaling}\fa, \st\  is kept fixed while varying \mup\  and, in Fig.~\ref{scaling}\fb, vice versa. It is clear from these graphs that \nev\  $\propto$ \ncl$^{\nu(\mu, \lambda)}$ and this is explicitly shown in Fig.~\ref{scaling}\fc, where all the graphs have been collapsed onto a single curve through $\nu = \mu^{-0.62} \times (1-\lambda)^{-0.2}$.}
\label{scaling}  
\end{figure*}  
In our analyses, we will be considering dust size distributions such that the mean dust size is comparable to the filter pore size. This turns out to be the regime of interest in which one can understand how filter and dust properties lead to the time evolution of the clogging process of the filter. Our studies show that dust size distributions can affect the details of the clogging process. 
 
{\it The Model System:}~~The present study is a 1$D$, mean-field-like, analysis. Dust grains are irregular in shape and come in all sizes. In our 1$D$ model, we only consider the size of the dust grains, i.e. effectively consider rod shapes. Consideration of all sizes in 1$D$ is accomplished by assuming a Gaussian distribution of dust sizes. Our studies suggest that an exponential tail to the distribution or a Lorentzian distribution {\it will} affect the rate at which filters clog. However, the major qualitative features remain the same for the different distributions considered and, therefore, Gaussian distribution will form the basis for our study.

Randomly sized rods from the distribution, assumed to be aligned vertically for easier geometric conceptualization, are allowed to impinge at random positions on a vertical 1$D$ filter with vertical slits or ``pores". The walls between the pores are considered to be of negligible width. Rods of all sizes travel at fixed speeds towards the filter. We assume that our time steps are sufficiently fine and/or the air is sufficiently dilute in dust content such that a single rod reaching the filter is a reasonable approximation. This justifies measurement of time, $t$, in terms of {\em particle release events} at `unit' time intervals of  $\Delta t$; $n^{{\mbox \em th}}$ particle is released at $t = n \times \Delta t$. Not all particles contribute to clogging. We will only count clogging events, and measure clogging rates, with respect to complete clogging of a pore. The number of clogged pores at any time is denoted as \ncl. One would expect \ncl\  to initially grow rapidly in time. But as time progresses, it will become progressively difficult for the rods to find the next available pore to clog. Thus, the temporal growth of \ncl\  will  eventually saturate. To best describe the manner in which \ncl\  grows, we define a new parameter to characterize the instant at   
which the $N^{{\mbox \em th}}$ clogging event happens and call it $t^{clog}_{N}$. This also leads us to the parameter $\tau$ ($\equiv t^{clog}_N - t^{clog}_{N-1}$), which is a {\em waiting time} between the $(N-1)^{{\mbox \em th}}$ and $N^{{\mbox \em th}}$ clogging. As time progresses and the filter gets progressively clogged, we expect the average value of $\tau$ to grow in magnitude. 
 
The mean of rod size, $r$, is referred to as $\mu$ and is kept smaller than the filter pore size, $s$. The standard deviation of rod sizes is taken to be unity. It is to be understood, however, that the variance {\em will} affect filter life, filtering efficiency etc. and will be discussed later. We study cases of filters with fixed pore sizes, as well as filters with a Gaussian ditsribution of pore sizes; filter pore size is initially kept fixed in the results reported below, and is to be considered so unless otherwise mentioned.  Ideally, when \szp\ $< s$, rods go through open pores, and, when \szp\  $>s$, they block open pores while, also, partially blocking adjacent pores by the amount by which $r > s$. However, we allow a rod with \szp\ slightly $< s$ to stick to a pore; this models the various sticking tendencies induced by diverse physical mechanisms, and also generates the well documented `bridge formation'~\cite{wata}, where several particles link together to arch over a pore. For this, we define a {\em sticking threshold}, $0 \le $ \st\ $\le 1$, which prescribes the rod length that can block an open pore as a fraction of $s$. If \st\ $= 1$, only rods with $r \ge s$ can block a pore, and hence this is the `non-stick' limit. If \st\ $< 1$, particles with $r < \lambda s$ will pass through the pore, without sticking. Typically, $0.5 \le $ \st\ $\le 0.9$ in our calculations, \mup\ is set arbitrarily equal to 5 units, and \szf\ to 9 units.

 We also allow for the formation of ``cakes'', another commonly observed phenomenon (see, for example, \cite{wata}),  with the additional deposition of rods on top of previously deposited rods, which may not contribute to additional blocking unless they extend over adjacent open pores. Rods are not assumed to hit the barrier between pores with any significant probability and get trapped by the same; this eliminates the need to consider the partial blocking of adjacent pores by a rod that drapes over the barrier. Further, adjacent pores are considered only to one side (in the direction of increasing array index) for convenience; these effects do not compromise the validity of the final results. In the following, we present our results on the clogging process of the 1$D$ filter. It may be noted that a 2$D$ filter can be thought of as an array of adjacent 1$D$ filters. In the limit of rods being incident at the same rate on each 1$D$ `unit' filter, the results of our present study are expected to remain applicable in the 2$D$ case.

{\em Results:} We consider a filter with \nzero\ pores at time $t = 0$; \nzero, in the following, is kept at $10^5$ for adequate statistical robustness. In Fig.~\ref{scaling}, we plot the time taken for the N$^{th}$ clogging event, \nev,  versus the number of clogged pores, \ncl, for various choices of the two control parameters, \mup\ and \st, in \lglg\ mode. In Fig.~\ref{scaling}\fa, \st\ is held fixed while varying \mup\  and, in Fig.~\ref{scaling}\fb, \mup\ is held fixed while \st\ is varied. All the cases are characterized by two different regimes of power law growth where, firstly, there is a smooth increase in \nev\  with \ncl, followed by an abrupt changeover to a region of infinite slope; the latter marks the divergent times, on the average, required to clog the final few pores, which are scattered randomly on the 1$D$ filter. Clearly, the interesting physics is in the first region, and we see that \nev\ $\propto$ \ncl$^\nu$, where $\nu = \nu($\mup, \st). The scaling pattern is made explicit in Fig.~\ref{scaling}\fc, where we have collapsed all the various graphs for the different \mup's and \st's onto a single graph, and it emerges that 
\begin{equation} 
\nu(\mu, \lambda) = \mu^{\alpha_1} \times (1.0-\lambda)^{\alpha_2}. 
\end{equation} 
where $\alpha_1 = -0.62$ and $\alpha_2 = -0.2$ for this case of Gaussian distribution of rod sizes impinging on a filter with fixed pore size. We found that these exponents are somewhat ``fragile'' and can vary depending on the nature of the rod and filter pore size distributions. For example, with an exponential distribution of rod sizes and a filter of fixed pore size, the exponents are $\alpha_1 = -0.2$ and $\alpha_2 = -0.1$. Nevertheless, the scaling behavior is robust.
 
\begin{figure}  
\centering  
\includegraphics[scale=0.43]{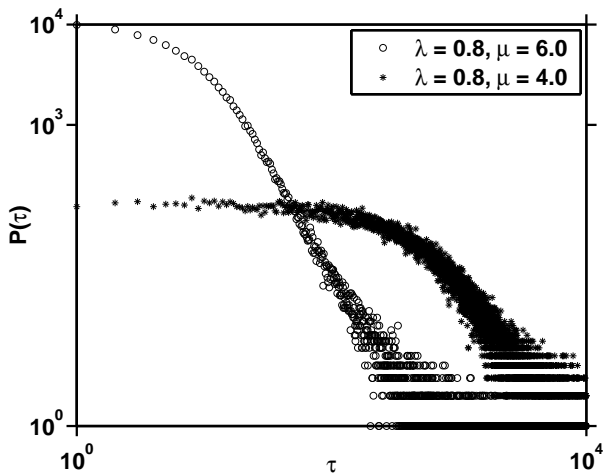}  
\caption{A plot of the distribution of waiting times, \dtw, against \tw, in a \lglg\  plot, shows a plateau region preceding the decay region when \mup\ is reduced; the graphs are with fixed pore size, \szf\ $= 9.0$. A similar plateau region results if \st\ is reduced, keeping \mup\  fixed. This indicates the enhanced role for smaller particles in the distribution, which are characterized by smaller \tw's, through the \inc\ mechanism.}  
\label{tautimes}  
\end{figure}  
Figs.~\ref{scaling}\fa-\fb\  reveal that the time required to clog all the pores tend to grow dramatically when \ncl\  becomes sufficiently large. We recall that the time period between successive cloggings is $\tau$. In Fig.~\ref{tautimes}, we present data of the distribution of $\tau$, $P(\tau)$, versus $\tau$, for two typical cases; the plots are in \lglg\ mode and only \tw's upto $10^4$ are shown. In one case, \dtw\ decays smoothly while, in the other case, the decay is preceded by a plateau region where the \tw's have the same density of occurrence. In general, it was observed that the plateau region increases with relative decrease of \mup\ with respect to \szf. Also, the same effect is generated if \st\ is decreased. In both these cases, we note that the smaller particles have an enhanced role to play in the clogging process and the plateau region is increased due to the phenomenon that we call \inc\ (IC). 
 
One can see that the role of the {\em big} ($r \geq \lambda s$) rods is necessary for the initiation of the clogging process; note that these come from the tail of the distribution, have smaller probabilities and, hence, larger \tw's associated with them. Once the process has been initiated, the smaller rods can contribute to additional clogging through the IC process, i.e. in the next stage, only a rod of size bigger than the \st\  fraction of the {\em remaining} open portion of the pore is required to further block the pore and so on. These smaller rods have larger probabilities and smaller \tw's associated with them. Also, if, at any stage, the rod that sticks on is big enough to protrude into the adjacent pore, it can reduce the size of that pore, which initiates IC in this adjacent pore. While the {\em big} rods contribute to clogging when IC is active, their role is especially significant in the initial stages of the clogging process. IC  increases the proportion of small \tw's and, as a corollary, when the role of smaller particles is enhanced either by reduction in \mup\ or by reduction in \st, IC becomes a more active mechanism. It is to be emphasized here that the random nature of the clogging events mandates non-zero probability of obtaining small \tw's even in the final stages of the clogging process; indeed, this has been clearly seen in plots of \tw's versus \ncl\ ~\cite{kmsen}.

\begin{figure*}
\centering  
\includegraphics[scale=0.80]{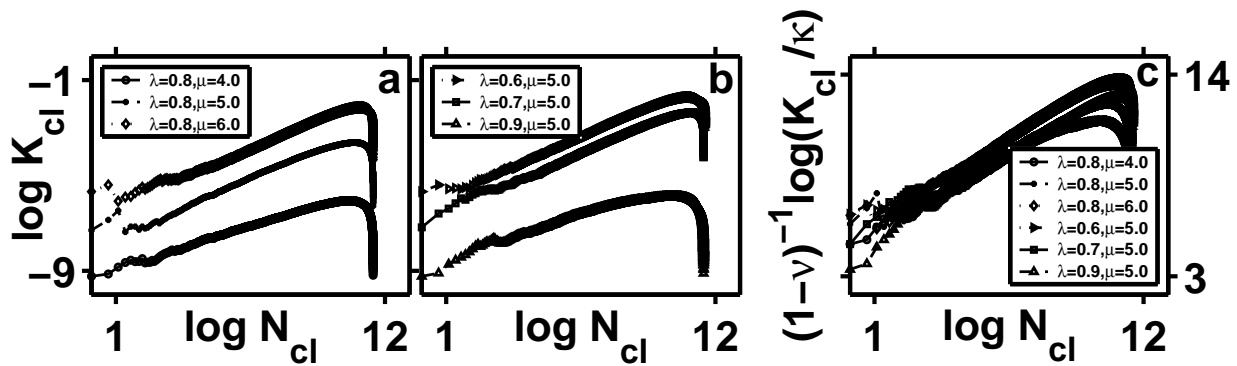}  
\caption{Plots of \kcl\ ($\equiv N_{cl}/$\nev; cf. Fig.~\ref{scaling}) versus \ncl, in \lglg\  mode; in Fig.~\ref{clgrate}\fa, \st\  is kept fixed while varying \mup\  and, in Fig.~\ref{clgrate}\fb, vice versa. A scaling relationship akin to Fig.~\ref{scaling}\fc\ may be suspected; this is indeed so, as shown by Fig.~\ref{clgrate}\fc, where all the graphs have been collapsed onto a single curve through proper scaling, with $\kappa = \exp(f(\mu, \lambda))$, $f(\mu, \lambda) = -16.3 + 1.04 \mu + 4.66 \lambda$.}  
\label{clgrate}  
\end{figure*}  
At the macroscopic level, it is of interest to probe the average rate of clogging at the time of clogging of the $N^{{\mbox \em th}}$ pore. We define this average rate of clogging via \kcl\ $\equiv N_{cl}/$\nev. This quantity is shown in Fig.~\ref{clgrate}, plotted against \ncl, for various choices of control parameters, \mup\ and \st. In Fig.~\ref{clgrate}\fa, \st\ is held fixed and \mup\ is varied while, in Fig.~\ref{clgrate}\fb, \mup\ is held fixed and \st\ is varied. The data suggests two different power-law growth regimes of $K_{cl}$ as a function of $N_{cl}$. There is a smooth increase in \kcl\ in the early stages followed by a steep decrease in the rate when there are only a few pores left to be clogged; the former regime corresponds to the functional period in the life of the filter. It turns out that the plots in Fig.~\ref{clgrate}\fa\  and Fig.~\ref{clgrate}\fb\  can be unified, as shown in Fig.~\ref{clgrate}\fc. The scaling relation used on the data in Fig.~\ref{clgrate}\fa\  and Fig.~\ref{clgrate}\fb\  turns out to be
\begin{equation} 
K_{cl} = \kappa \times N_{cl}^{1-\nu}.
\end{equation} 
where $\kappa = \exp (f(\mu, \lambda))$, and $f(\mu, \lambda) = -16.3 + 1.04 \mu + 4.66 \lambda$. The relationship with respect to \ncl\ is to be expected given the definition of \kcl\ and Eq.~(1). However, further complexity of the scaling relationship via the prefactor $\kappa$ results because we are considering a rate quantity in a problem with discontinuous kinetics.
 
\begin{figure}  
\centering  
\includegraphics[scale=0.43]{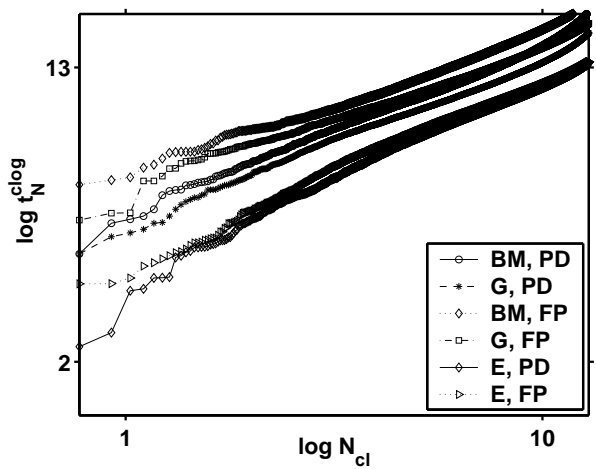}  
\caption{Effect, on clogging rates, of different rod size distributions, with and without filter pore size distribution are compared. The notation in the legend is: BM-- Bimodal Gaussian (two modes being at 5.0 and 3.0), G-- Gaussian (mean at 5.0), E-- Exponential (mean at 5.0); FP-- fixed pore size (\szf\ = 9.0), PD-- distribution of pore sizes with mean at 9.0. It is seen that faster clogging takes place with a pore size distribution in general, and bimodal distribution is the slowest and exponential distribution fastest in clogging the filter.}
\label{cmprates}  
\end{figure}  
It is easy to calculate (see, for example,~\cite{schaum}), based on the probabilities for particles of size $r > k\sigma$, where $\sigma$ is the standard deviation of rod size distribution and $k$ any positive value, the probability of finding a rod of size $r \geq s$ as against finding a rod of size $r \geq \lambda s$; it turns out that the latter is greater by three orders of magnitude if $\lambda = 0.8$, with $\mu = 5.0$, $\sigma = 1.0$ and fixed pore size $s = 9.0$~\cite{kmsen}. Differences in probability translate to comparable differences in \nev. Clearly, variance of the distribution also controls the rate of clogging. Similar differences in probability can be achieved by varying \mup\ with respect to \szf. 

If we introduce a distribution of pore sizes around a mean value, instead of having a fixed pore size, \nev\  decreases; in this comparison, the same value of the fixed pore size was assigned for the mean. If we introduce a second mode at a lower value in a corresponding unimodal Gaussian distribution, \nev\ increases because of the enhanced proportion of smaller sized rods; in this comparison, a value of 3.0 was used for the second mode with the first mode at 5.0. An exponential distribution, which has a higher proportion of bigger rods than a Gaussian, leads to faster clogging of the pores. These results are summarized in Fig.~\ref{cmprates} in a \lglg\ plot of \nev\ {\em versus} \ncl. A corresponding ordering, with respect to the distributions, can be observed in plots of \kcl\  {\em versus} \ncl\ plots; note that larger times translate to lower \kcl\ values in this case. 
 
{\em Concluding Remarks:}~~ Our studies indicate that adjusting the filter pore size (or mean of the same in case of a pore size distribution) to be slightly higher than the mean of the dust particle distribution is an important step in obtaining efficient filtering as well as longer filter lifetimes. The variance as well as the type of distribution of the filter and dust particles are also important determinants. The scaling law that we propose in the relationship of \nev\ and \ncl, with respect to the control parameters \mup\ and \st, can play an important role in suitably choosing the filter parameters. Even though the scaling exponent $\nu$ is somewhat ``fragile'' with respect to different distributions of dust particles, the fact that its magnitude is only altered by a factor of the order of 2, rather than by an order of magnitude, in going from a Gaussian distribution to an exponential distribution, is very encouraging. A key parameter for further analysis is the resistance to the flow caused by filter clogging which requires modelling of fluid flow; the relationship of cakes to the model parameters would also be interesting to study. The present study is expected to be applicable to any problem where small particles can lead to clogging of any fluid flow. The model may have broad applicability with respect to problems where the concept of \inc\ can be applied.

{\it Acknowledgements}: The research has been supported by NSF CMS0070055. We thank Adam Sokolow and Masami Nakagawa for their interest in this study. 
 

\end{document}